\documentclass[sigconf]{acmart}

\AtBeginDocument{%
  }

\setcopyright{acmlicensed}
\copyrightyear{2026}
\acmYear{2026}
\acmDOI{10.1145/XXXXXX.XXXXXX} %
\acmConference[ACM MM '26]{ACM International Conference on Multimedia}{November 10--14,
  2026}{Rio de Janeiro, Brazil}
\acmISBN{978-1-4503-XXXX-X/26/06}

\usepackage{pifont} % for \cmark and \xmark
\usepackage{multirow}

\usepackage{amssymb}  
\usepackage{makecell}
\usepackage[table]{xcolor}
\usepackage{booktabs}
\usepackage{bbding}
\usepackage[most]{tcolorbox}

\begin{document}

\title{QKVQA: Question-Focused Filtering for Knowledge-based VQA}

\author{Wei Ye}
\email{weiye@hust.edu.cn}
\affiliation{%
  \institution{Huazhong University of Science and Technology}
  \city{Wuhan}
  \country{China}
}

\author{Yixin Su}
\email{suyixin@hust.edu.cn}
\affiliation{%
  \institution{Huazhong University of Science and Technology}
  \city{Wuhan}
  \country{China}
}

\author{Yueguo Chen}
\email{chenyueguo@gmail.com}
\affiliation{%
  \institution{Renmin University of China}
  \city{Beijing}
  \country{China}
}

\author{Longxiang Gao}
\email{gaolx@sdas.org}
\affiliation{%
  \institution{Qilu University of Technology (Shandong Academy of Sciences)}
  \city{Jinan}
  \country{China}
}

\author{Jianjun Li}
\email{jianjunli@hust.edu.cn}
\affiliation{%
  \institution{Huazhong University of Science and Technology}
  \city{Wuhan}
  \country{China}
}

\author{Ruixuan Li}
\email{rxli@hust.edu.cn}
\affiliation{%
  \institution{Huazhong University of Science and Technology}
  \city{Wuhan}
  \country{China}
}

\author{Rui Zhang}
\authornote{Corresponding Author.}
\email{rayteam@yeah.net}
\affiliation{%
  \institution{Huazhong University of Science and Technology}
  \city{Wuhan}
  \country{China}
}

\renewcommand{\shortauthors}{Ye et al.}

\settopmatter{printacmref=false}
\renewcommand\footnotetextcopyrightpermission[1]{}

\begin{abstract}
Visual Question Answering (VQA) is the task of answering questions based on image content. Building upon this, Knowledge-Based VQA (KB-VQA) requires models to answer questions that depend on external knowledge beyond the visual content of an image. In such settings, effective knowledge filtering is essential for achieving high question answering accuracy. Typical filtering methods suffer from two issues: they fail to focus on parts relevant to the question during candidate section encoding, and they use similarity metrics to locate a section from a single article, resulting in information limitation. To address these issues, this paper proposes a question-focused, cross-article filtering method. Specifically, we design a trainable Question-Focused Filter (QFF) and a Chunk-based Dynamic Cross-Article Selection module (CDA). This approach maintains inference time comparable to the optimal method with the shorter context length, efficiently obtaining high-quality filtered knowledge. The accuracy outperforms current state-of-the-art methods by 3.2 and 2.2 percentage points on Encyclopedic-VQA and InfoSeek, respectively. The code is publicly available at: https://anonymous.4open.science/r/QKVQA.
\end{abstract}




\received{20 February 2007}
\received[revised]{12 March 2009}
\received[accepted]{5 June 2009}

\maketitle

\section{Introduction}

\begin{figure}[h]
  \centering
  \includegraphics[width=0.9\linewidth]{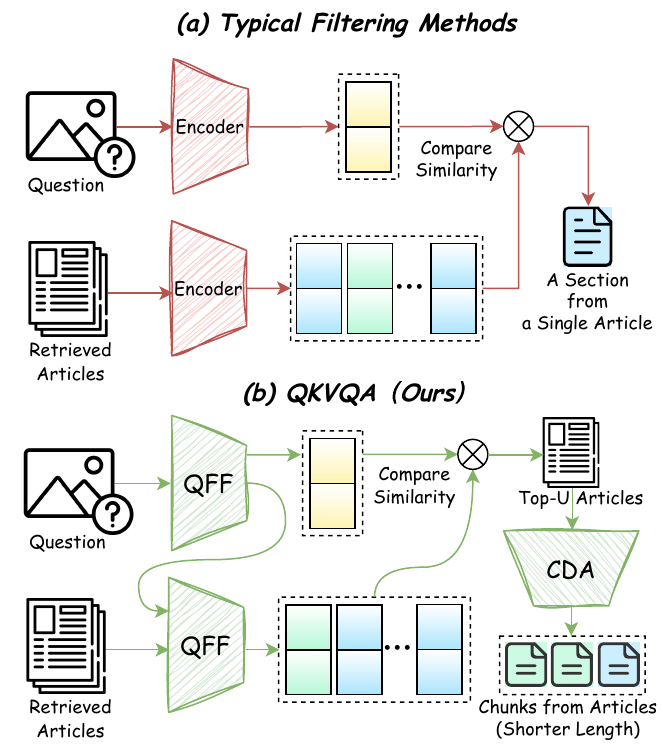}
  \vspace{-0.4cm}

  \caption{Comparison between typical filtering methods and our proposed approach.}
  \label{fig:intro}
\vspace{-0.3cm}
\end{figure}
Visual Question Answering (VQA) is the task of answering questions about visual content such as images or videos. With the advancement of Multimodal Large Language Models (MLLMs)~\cite{zhang2025comprehensive,bai2025qwen2,cocchi2025llava,liu2023improved}, VQA has seen rapid progress. While MLLMs perform well on many VQA tasks~\cite{xiao2025flair,wang2025mira,tschannen2025siglip,zi2025rsvlm,yang2025magic,kim2025x,li2023blip}, visual content alone is often insufficient for questions requiring external knowledge. Knowledge-based VQA (KB-VQA) overcomes this by retrieving external knowledge from knowledge bases~\cite{sravanthi2025rg,chen2023can,mensink2023encyclopedic,marino2021krisp}.

The Retrieval-Augmented Generation (RAG) framework offers a viable solution for KB-VQA~\cite{lin2024preflmr,si2023combo}, which typically involves three stages: retrieval, filtering, and generation~\cite{hong2025knowledge}. Effective knowledge filtering is crucial for VQA accuracy. As shown in Figure~\ref{fig:intro}(a), typical methods~\cite{yang2025omgm,yan2024echosight} face two main issues.

\begin{figure}[t]
  \centering
  \includegraphics[width=\linewidth]{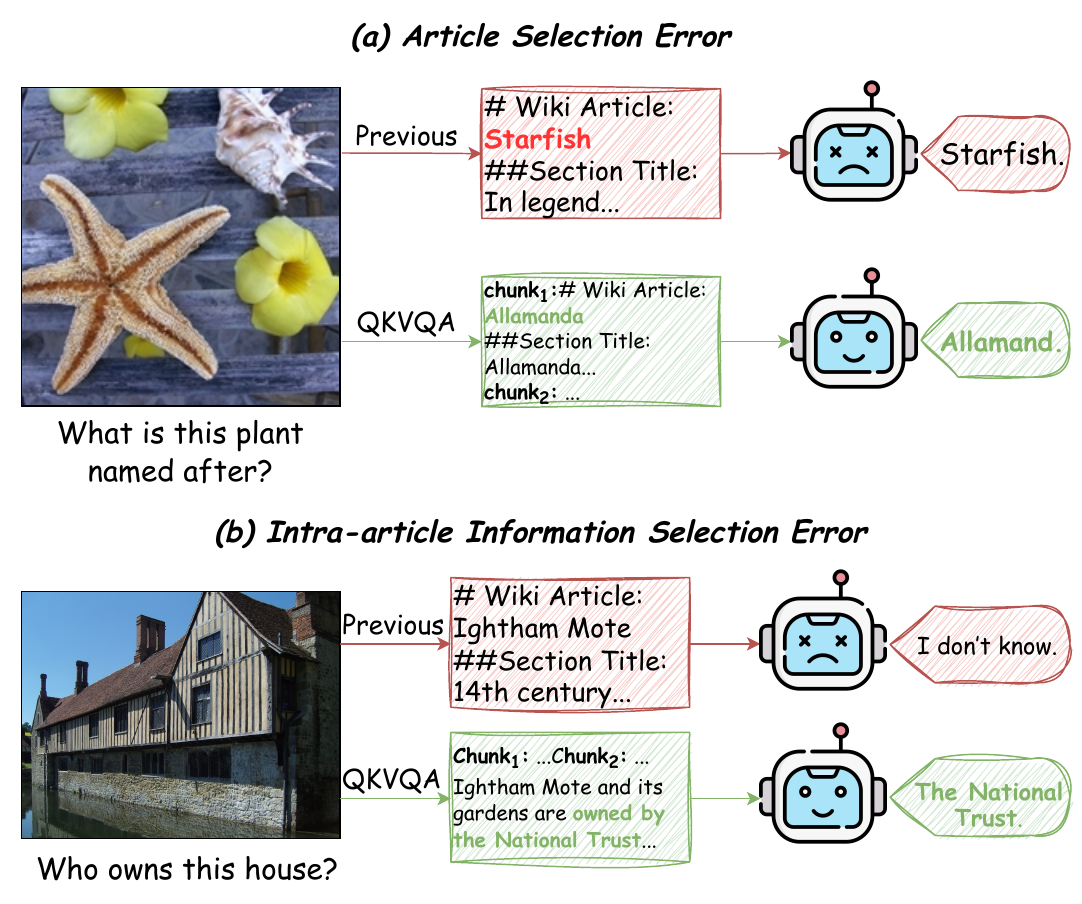}
  \vspace{-0.5cm}
  \caption{Concrete examples of the two types of errors.}
  \label{fig:intro2}
\vspace{-0.4cm}
\end{figure}

First, typical methods independently encode the question and candidate sections, lacking the ability to focus on parts relevant to the question during candidate section encoding. This results in a low similarity ranking between the correct article and the question, ultimately leading to article selection errors. As illustrated in Figure~\ref{fig:intro2}(a), when presented with an image containing multiple objects (e.g., a starfish, a seashell, and a plant) and asked, ``What is this plant named after?'', typical methods independently encode sections from candidate knowledge articles and select the single article with the highest similarity score (e.g., the article about the starfish), ultimately producing an incorrect answer. If, during candidate encoding, the model focuses specifically on the ``plant''-related parts of the question, this issue can be mitigated.

Second, they suffer from information limitation. At the article level, typical methods are confined to filtering from a single article, which may cause them to miss the correct article, resulting in the article selection error described above. At the intra-article level, even when the correct article is selected, these methods may still fail to capture the specific textual fragments required to answer the question. We refer to this as the intra-article selection error. As shown in Figure~\ref{fig:intro2}(b), for the question ``Who currently owns Ightham Mote?'', typical filtering methods may retrieve lengthy sections describing historical owners, yet fail to include the key section that indicates the current owner, ultimately leading to an incorrect answer.

Although typical methods can mitigate the problem of information limitation by expanding the selection range from top-1 to top-k at both the article and section levels, this introduces excessive context length and inference time for the subsequent Large Language Model answer generator.

Try to alleviate the two issues mentioned above, recent studies have explored the use of MLLMs for filtering. Such models are capable of focusing on parts relevant to the question during the filtering process and achieve cross-article content consolidation~\cite{hong2025knowledge,compagnoni2025reag,cocchi2025augmenting,ling2025mmkb,tian2025core,zhang2024mr,caffagni2024wiki}. However, MLLM-based methods typically require inputting complete multiple articles~\cite{cocchi2025augmenting} or the majority of sections from multiple articles~\cite{hong2025knowledge}. This, similar to directly expanding the selection range of typical methods, not only leads to excessively long context lengths and significantly increased inference time but also introduces excessive noisy information that can interfere with model judgment. For these reasons, such methods remain impractical for applications.

To maintain the context length and inference time of typical methods while leveraging the advantages of the aforementioned MLLM-based filtering methods, this paper proposes a question-focused filtering approach (Figure~\ref{fig:intro}(b)). This goal is achieved through the following two key designs:

(1) Trainable Question-Focused Filter (QFF): Built upon the Q-Former architecture, the QFF first encodes the question using its learnable queries and retains their hidden states at each encoder layer. When encoding candidate knowledge, a cross-attention mechanism is introduced at each layer of the encoder to deeply integrate the hidden states, which encapsulate the question semantics, into the queries of the current layer. This progressively infuses question-focused semantic cues into the knowledge encoding process. This mechanism enhances the perception of semantic relevance and effectively reduces article selection errors.

(2) Chunk-based Dynamic Cross-Article Selection Module (CDA): This module transforms traditional coarse-grained section-level filtering into a fine-grained, cross-article dynamic selection strategy based on text chunks. At the article level, we design a dynamic cross-article selection module based on QFF scores to control the overhead introduced by processing multiple articles. Specifically, within a fixed scope, articles are filtered by comparing the score difference between each article and the highest-scoring article against a predefined threshold, determining which articles are retained for further filtering. At the intra-article level, inspired by chunk-level retrieval in text-based RAG~\cite{li2025raidx,luo2025ma,zhang2025comprehensive}, we refine the filtering granularity from sections to text chunks. Furthermore, considering the dependency of KB-VQA tasks on fine-grained information and meta-information (e.g., ``article title'', ``section title''), we split overly long sections into text chunks that do not exceed a preset length while preserving sentence integrity, and retain their original meta-information for each chunk. This strategy not only enables access to more fine-grained information but also effectively reduces context length, thereby lowering noise interference and decreasing inference time costs. We first filter the articles to be retained for further processing, then divide them into the aforementioned text chunks and perform further ranking. This module enables the model to precisely locate required information from multiple articles, thereby effectively alleviating the issue of information limitation, and consequently mitigating both article selection errors and intra-article information selection errors.

The main contributions of this paper are summarized as follows:

\begin{itemize}
    \item We propose a trainable, question-focused filter that embeds question semantics into the candidate section encoding, enhancing the filter's ability to focus on parts relevant to the question during filtering and alleviating article selection errors.
    \item We design a fine-grained, chunk-based dynamic cross-article selection module that supports fine-grained information filtering and cross-article collaboration, effectively alleviating article selection and intra-article information selection errors.
    \item The accuracy on two challenging KB-VQA benchmarks, InfoSeek and Encyclopedic-VQA, outperforms current state-of-the-art models by 3.2\% and 2.2\%, respectively, while maintaining inference time comparable to the optimal method and utilizing shorter context lengths.
\end{itemize}

\section{Related Work}
\subsection{Knowledge-based Visual Question Answering}
Knowledge-based Visual Question Answering (KB-VQA) requires models to answer questions that depend on external knowledge beyond the visual content of an image. Early datasets~\cite{schwenk2022okvqa,marino2019ok,shah2019kvqa} primarily focused on commonsense knowledge. With the advancement of Multimodal Large Language Models (MLLMs), such data have become insufficient for knowledge-intensive scenarios. Recent datasets such as Encyclopedic-VQA~\cite{mensink2023encyclopedic} and InfoSeek~\cite{chen2023can} emphasize fine-grained article reasoning, demanding stronger filtering capabilities. Currently, Retrieval-Augmented Generation (RAG)~\cite{lewis2020retrieval} has become the mainstream technical framework in this field.

\subsection{Filtering Mechanisms in KB-VQA}

Filtering mechanisms play a critical denoising role in the RAG pipeline. Typical filtering methods~\cite{yang2025omgm,yan2024echosight} primarily operate by independently encoding the question and candidate knowledge and then comparing their similarity, lacking the ability to focus on parts relevant to the question during encoding. Moreover, to control noise and computational costs, they are typically limited to a section of a single article. These issues easily lead to article selection and intra-article information selection errors.

Recent research has begun to explore filtering approaches utilizing Multimodal Large Language Models (MLLMs)~\cite{hong2025knowledge,compagnoni2025reag,cocchi2025augmenting,ling2025mmkb,tian2025core,zhang2024mr,caffagni2024wiki}. These methods can better focus on parts relevant to the question during the filtering process and leverage the reasoning capabilities of large models to integrate information across multiple articles. However, MLLM-based methods typically require feeding entire articles or their major sections into the model for processing. For instance, ReflectiVA~\cite{cocchi2025augmenting} uses an MLLM to analyze, filter, and integrate all passages from the top-5 retrieved articles, while VLM-PRF~\cite{hong2025knowledge} selects and integrates the top-3 most relevant passages from the top-5 articles. Such approaches not only lead to excessive context length and significantly increased inference time but also introduce excessive noisy information that can interfere with model judgment, thereby making such methods remain impractical for applications.

This paper combines the strengths of both types of approaches, proposing a question-focused filter and a chunk-based dynamic cross-article selection module, aiming to simultaneously alleviate article selection error and intra-article information selection error.

\begin{figure*}[h]
  \centering
  \includegraphics[width=\linewidth]{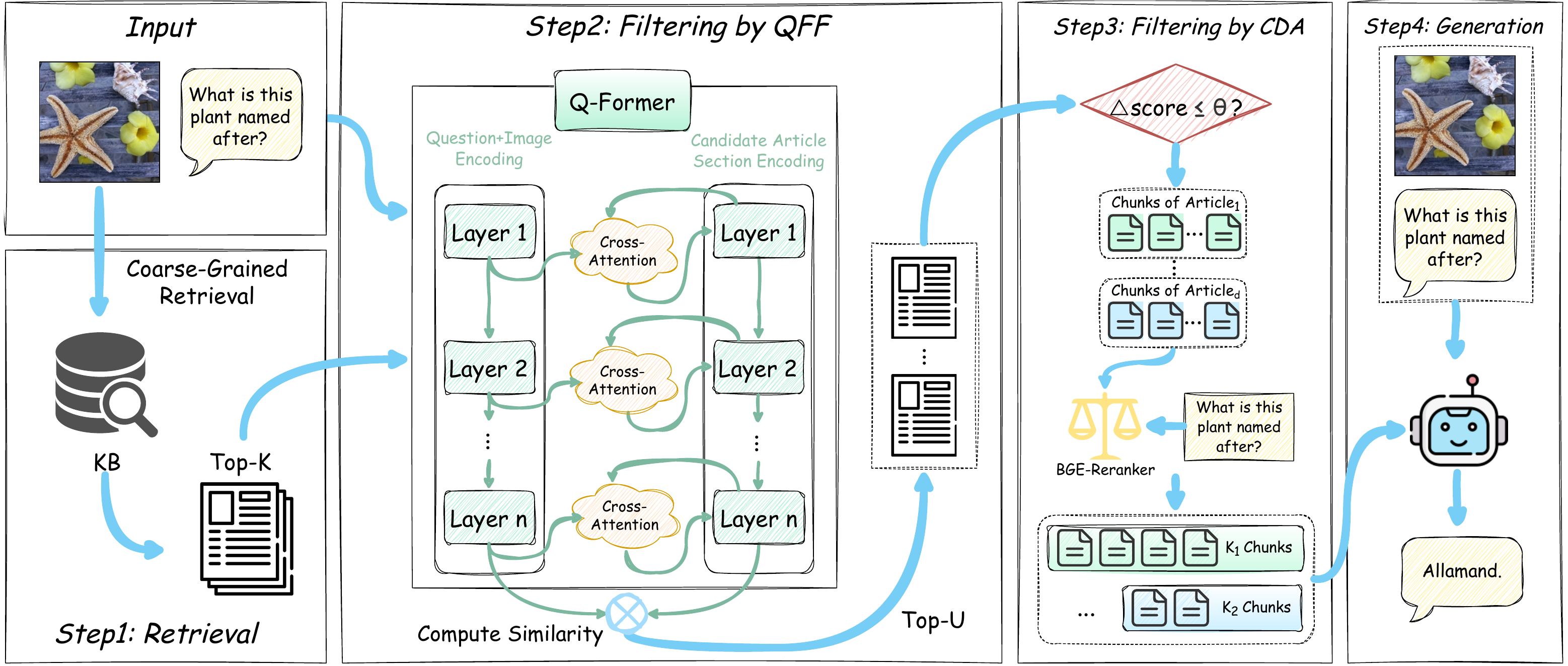}
  \vspace{-0.6cm}

  \caption{The overview of our complete pipeline.}
  \label{fig:overview}
\vspace{-0.4cm}

\end{figure*}

\section{Proposed Method}
\subsection{Retrieval}
We first employ a coarse-grained cross-modal retriever to fetch candidate articles from the knowledge base $\mathcal{KB}$ (Step 1 in Figure~\ref{fig:overview}). Given the query image $I_q$, we encode it with a visual encoder $E_v$ and the abstract $T_i$ of each article with a text encoder $E_t$. Their similarity is measured by cosine distance:
\begin{equation}
\mathrm{sim}^{\mathrm{ret}}_i = \frac{E_v(I_q) \cdot E_t(T_i)}{\|E_v(I_q)\|\,\|E_t(T_i)\|}.
\end{equation}
The top-$K$ articles form the candidate set:
\begin{equation}
\mathcal{E}_{\mathrm{top-}K} = \{ a_{(1)}, a_{(2)}, \dots, a_{(K)} \},
\label{TOPK}
\end{equation}
where $a_{(i)}$ denotes the article with the $i$-th highest similarity score.

Each candidate article is then segmented into structured sections:
\begin{equation}
s_i = \{(t_{ij}, p_{ij}, I_i)\},
\end{equation}
where $t_{ij}$ is the metadata (article and section titles), $p_{ij}$ is the passage content, and $I_i$ is the article image.

\subsection{Filtering by QFF}
\label{QFF}

Typical filtering methods often compute the relevance between a question and candidate sections independently, lacking the ability to focus on parts relevant to the question during candidate encoding. We propose the Question-Focused Filter (QFF), based on the Q-Former architecture, where a set of learnable queries extracts key multimodal information under the semantic guidance of the question (Step 2 in Figure~\ref{fig:overview}).

Specifically, the question $q$ and the image $I_q$ are encoded by the Q-Former to generate a fused question representation $Q\_tokens \in \mathbb{R}^{N \times d}$:
\begin{equation}
Q\_tokens = \text{QFF}(q, I_q \mid Queries),
\label{N}
\end{equation}
where $Queries \in \mathbb{R}^{N \times d}$ are learnable queries, $N$ is their number, and $d$ is the feature dimension. In the Q-Former with $n$ encoding layers, the states of the queries during the encoding process are denoted as $Queries_1, Queries_2, \ldots, Queries_n, Queries_{n+1}$, where $Queries_1$ is the initial state before the first layer, $Queries_{i+1}$ is the output state after the $i$-th layer for $i = 1, \ldots, n$, and after $n$ layers of encoding, the output $Q\_tokens$ is obtained, i.e., $Queries_{n+1} = Q\_tokens$.

When encoding candidate sections, we use the same Q-Former as used for encoding the question. We set the initial state of the queries for encoding sections as $F\_Queries_1 = Queries_1$. To make the queries focus on question-related information, in each encoding layer, we fuse the pre-encoding state $F\_Queries_i$ with the post-encoding state $Queries_{i+1}$ from the corresponding layer of question encoding via cross-attention:
\begin{equation}
F\_Queries_i = F\_Queries_i + \mathrm{Softmax}\left(\frac{QK^{\top}}{\sqrt{d}}\right)V,
\end{equation}
where $Q = F\_Queries_iW^Q$, $K = Queries_{i+1}W^K$, $V = Queries_{i+1}W^V$, and $W^Q, W^K, W^V \in \mathbb{R}^{d \times d}$ are learnable parameters. The fused $F\_Queries_i$ is then fed into the $i$-th encoding layer, producing the output state $F\_Queries_{i+1}$. This operation injects question semantics into the queries, enhancing their question-focused selection capability.

After $n$ layers of encoding, for the $j$-th section $s_{ij} = (t_{ij}, p_{ij}, I_i)$ of the $i$-th article containing $L_i$ sections, the encoded feature $h_{i,j}$ is:
\begin{equation}
h_{i,j} = F\_Queries_{n+1} = \text{QFF}\bigl(s_{ij} \mid F\_Queries\bigr), \quad j=1,\dots,L_i.
\end{equation}

Using a delayed interaction mechanism~\cite{khattab2020colbert}, we compute fine-grained relevance between the question and each section. For each token vector in $h_{i,j}$ and each token vector in $Q\_tokens$, we take the maximum cosine similarity. The section similarity is the average of these maximum values:

\begin{equation}
\mathrm{sim}^{\mathrm{QFF}}_{i,j} = \frac{1}{|h_{i,j}|} \sum_{t \in h_{i,j}} \max_{q \in Q\_tokens} \frac{q \cdot t}{\|q\|\|t\|}.
\end{equation}

The QFF score for an article is the highest similarity among all its sections:

\begin{equation}
\mathrm{sim}^{\mathrm{QFF}}_i = \max_{j=1,\dots,L_i} \mathrm{sim}^{\mathrm{QFF}}_{i,j}.
\end{equation}

Finally, we filter and rank articles by fusing the QFF score with the retrieval score via weighted combination:

\begin{equation}
\mathrm{sim}^{\mathrm{fused}}_i = \alpha \cdot \mathrm{sim}^{\mathrm{ret}}_i + (1-\alpha) \cdot \mathrm{sim}^{\mathrm{QFF}}_i,
\label{a}
\end{equation}
where $\alpha \in [0,1]$ balances the two scores. We select the top-U articles, generating the refined article list $\mathcal{E}_{\mathrm{QFF}}$.

\subsection{Training of QFF}
\label{Training of QFF}

The QFF is trained using contrastive learning to distinguish relevant sections from irrelevant ones. For each question-image pair $(q, I_q)$ and its corresponding positive section $s^+$, we sample $M$ negative sections $\{s^-_1, \dots, s^-_M\}$ from the top-$K$ articles retrieved.

Following the feature extraction and scoring pipeline described in Section~\ref{QFF}, $s^+$ is encoded by the QFF as $\mathrm{sim}^{\mathrm{QFF}}_+$, and the negative sections $\{s^-_1, \dots, s^-_M\}$ are encoded as $\{\mathrm{sim}^{\mathrm{QFF}}_{-,1}, \dots, \mathrm{sim}^{\mathrm{QFF}}_{-,M}\}$, respectively. We then compute:
\begin{equation}
\mathcal{L}_{\mathrm{QFF}} = -\log \frac{\exp\!\bigl(\mathrm{sim}^{\mathrm{QFF}}_+ / \tau\bigr)}
{\exp\!\bigl(\mathrm{sim}^{\mathrm{QFF}}_+ / \tau\bigr) + \sum_{m=1}^{M} \exp\!\bigl(\mathrm{sim}^{\mathrm{QFF}}_{-,m} / \tau\bigr)},
\label{tem}
\end{equation}
where $\tau > 0$ is a temperature parameter. This temperature‑scaled cross‑entropy loss increases the score gap between the positive section and the negative ones, improving the model's discriminative ability.

\subsection{Filtering by CDA}
\label{CDA}
To achieve finer-grained filtering within and across articles while reducing context length and controlling time consumption, we propose the Chunk‑based Dynamic Cross‑Article Selection module (CDA). CDA dynamically selects the most relevant text chunks from $\mathcal{E}_{\mathrm{QFF}}$.

We first take the top-$U$ articles from $\mathcal{E}_{\mathrm{QFF}}$ and compute the QFF score difference relative to the top-1 article:
\begin{equation}
\Delta_i = \mathrm{sim}^{\mathrm{QFF}}_{1} - \mathrm{sim}^{\mathrm{QFF}}_{i},\quad i=1,\dots,U.
\label{TOPU}
\end{equation}
This score difference reflects the confidence level of QFF in the top-1 selected article. Articles satisfying $\Delta_i \le \theta$ are retained, where $\theta$ is a preset threshold. The final number of selected articles is denoted as $D$, which corresponds to the total number of articles meeting the threshold condition and represents the top-$D$ articles with the highest similarity to the question in terms of multimodal features. Since that all sections within the same article share the same image $I_i$, we next perform focused filtering at the textual feature level within each article, aiming to identify evidence sentences.

The sections of each selected article are split into text chunks of a fixed maximum length $L$ while preserving sentence integrity. Sections shorter than $L$ or individual sentences exceeding $L$ remain as single chunks; longer sections are divided. Each chunk $c_{i,j'}$ retains its metadata $t_{i,j}$ (article and section titles) to preserve structural context. Here, the subscript $i$ denotes the article index, $j$ denotes the section index within that article, and $j'$ denotes the chunk index derived from section $j$.

We then use BGE-Reranker-v2-m3~\cite{chen-etal-2024-m3} as the reranker $R_t$ to compute the relevance between each chunk $c_{i,j'}$ and the textual content $q$ in the question:
\begin{equation}
\mathrm{sim}^{\mathrm{chunk}}_{i,j'} = R_t(q, c_{i,j'}).
\end{equation}

It is worth mentioning that OMGM~\cite{yang2025omgm} also employs BGE-Reranker-v2-m3, which allows for a fair comparison.

For each selected article, we directly filter based on the chunk scores $\mathrm{sim}^{\mathrm{chunk}}_{i,j'}$ computed by the reranker. To prioritize higher-ranked articles within a context length as short as possible, we allocate a larger chunk quota to the top article: we set $K_1 > K_2$, where the first article selects $K_1$ highest-scoring chunks, and each of the remaining $D-1$ articles selects $K_2$ highest-scoring chunks. The final set of selected chunks is:
\begin{equation}
\mathcal{C}_{\mathrm{selected}} = \bigcup_{i=1}^{D} \bigl\{ c_{i,(1)}, c_{i,(2)}, \dots, c_{i,(K_i)} \bigr\},
\label{k1k2}
\end{equation}
where $c_{i,(k)}$ denotes the $k$-th highest-scoring chunk in the $i$-th article, and $K_i$ takes the value $K_1$ (when $i=1$) or $K_2$ (when $i \geq 2$).

\subsection{Answer Generation}
\label{Answer Generation}
The selected chunk set $\mathcal{C}_{\mathrm{selected}}$ is concatenated into a coherent context string, which together with the original question $q$ and its corresponding image $I_q$ (or an empty input if the generator is an LLM) is fed into the generator $G$. The generator produces the final answer $A$:
\begin{equation}
A = G\bigl(q, I_q, \mathcal{C}_{\mathrm{selected}}\bigr).
\end{equation}

\section{Experiments}
\subsection{Datasets and Evaluation Metrics}
We use two challenging datasets: Encyclopedic-VQA (E-VQA)~\cite{mensink2023encyclopedic} and InfoSeek~\cite{chen2023can}. The E-VQA test set contains 4,750 single-hop and 1,000 two-hop questions. Following prior works~\cite{yang2025omgm,yan2024echosight,cocchi2025augmenting}, when involving recall computation, including Table~\ref{tab:two_error}, Table~\ref{tab:ablation_retrieval_scope}, and Figure~\ref{fig:alpha_lambda}, we conduct experiments using single-hop questions for those parts. Beyond that, we report the complete results including two-hop questions. Also following the prior works, we use the same 71,335 validation samples for InfoSeek. Additionally, we adopt the same knowledge base settings.

For VQA evaluation, we adopt each dataset’s official metrics: BEM score for E-VQA~\cite{zhang2019bertscore} and both VQA accuracy and relaxed accuracy for InfoSeek~\cite{antol2015vqa,methani2020plotqa}. Retrieval performance is measured via Recall@K. More details on the datasets and their evaluation methods can be found in the Appendix.

\subsection{Implementation Details}
\subsubsection{Retrieval Details}
Following~\cite{yang2025omgm,cocchi2025augmenting,yan2024echosight}, we encode the query image with EVA‑CLIP‑8B~\cite{sun2024eva} for coarse‑grained retrieval, and perform large‑scale search via the faiss library from OMGM~\cite{yang2025omgm}, retaining the top‑$K$ ($K=20$) results as in prior work (Eq.~\ref{TOPK}).

\subsubsection{Filter Training and Inference}
The encoder is initialized with Q‑Former weights from the LAVIS library~\cite{li-etal-2023-lavis}, using the first 32 embedding vectors as the multimodal fusion feature matrix. Due to the lack of evidence section annotations in InfoSeek, the encoder is trained on E‑VQA and evaluated on both datasets to test generalization.

Training employs hard-negative sampling with 16 image-section pairs per sample (1 positive, 15 negative). For fair comparison, we adopt the same hard-negative sampling strategy as OMGM~\cite{yang2025omgm}.

We train for one epoch on 190K randomly sampled E-VQA samples with a learning rate of $1\times10^{-5}$ and a batch size of 4, completing in approximately 14 hours on two NVIDIA L40 GPUs. The inference hyperparameters are set as follows: similarity fusion weight $\alpha = 0.9$ (Eq.~\ref{a}), temperature $\tau = 0.07$ (Eq.~\ref{tem}), article count $U = 2$ (Eq.~\ref{TOPU}), threshold $\theta = 0.01$ (Eq.~\ref{TOPU}), and chunk quotas $K_1 = 4$, $K_2 = 2$ (Eq.~\ref{k1k2}). Additionally, to control context length (tokens), the chunk size described in Section~\ref{CDA} is set to 90 tokens of the generator's tokenizer for InfoSeek and 60 tokens for E-VQA.

\begin{table*}[ht!]
\caption{VQA accuracy comparison with the baselines. Gen. FT indicates whether the generator of the method was fine-tuned. Best in bold, second-best underlined. The markers $\dagger$ and $\ddagger$ represent our reproductions and using the fine-tune generator supported by ReflectiVA, respectively.}
\vspace{-2mm}
\label{tab:vqa_main_results}
\centering
\scalebox{1.0}{
\renewcommand{\arraystretch}{1.0}
\resizebox{\linewidth}{!}{
\begin{tabular}{l|c|c|cc|ccc}
\toprule
\multirow{2}{*}{\textbf{Method}} & \multirow{2}{*}{\textbf{Generator}} & \multirow{2}{*}{\textbf{Gen. FT}} & \multicolumn{2}{c|}{\textbf{E-VQA}} & \multicolumn{3}{c}{\textbf{InfoSeek}} \\
\cline{4-8}
 & & & \textbf{Single-Hop} & \textbf{All} & \textbf{Unseen-Q} & \textbf{Unseen-E} & \textbf{Overall} \\
\midrule
LLaMA-3-8B~\cite{dubey2024llama} & LLaMA-3-8B & -- & 16.3 & 17.3 & 1.5 & 0.0 & 0.0 \\
LLaMA-3.1-8B~\cite{dubey2024llama} & LLaMA-3.1-8B & -- & 16.5 & 16.6 & 2.1 & 0.0 & 0.0 \\
GPT-4~\cite{achiam2023gpt} & GPT-4 & -- & 21.9 & 23.4 & 7.3 & 5.0 & 5.9 \\

Qwen2.5-VL-7B~\cite{bai2025qwen2} & Qwen2.5-VL-7B & -- & 23.6 & 23.2 & 22.8 & 24.1 & 23.7 \\
GPT-4V~\cite{achiam2023gpt} & GPT-4V & -- & 26.9 & 28.1 & 15.0 & 14.3 & 14.6 \\
\midrule

VLM-PRF~\cite{hong2025knowledge} & LLaMA-3.1-8B & $\checkmark$ & 36.3 & 35.5 & 41.3 & \underline{40.6} & \underline{40.8} \\
mKG-RAG~\cite{yuan2025mkg} & LLaMA-3.1-8B & $\checkmark$ & 38.4 & 36.3 & \underline{41.4} & 39.6 & 40.5 \\
ReflectiVA~\cite{cocchi2025augmenting} & LLaMA-3.1-8B & $\checkmark$ & 35.5 & 35.5 & 40.4 & 39.8 & 40.1 \\
OMGM\textsuperscript{\small$\ddagger$}~\cite{yang2025omgm} & LLaMA-3.1-8B  & $\checkmark$ & \underline{46.5} & \underline{44.1} & 41.0& 39.6& 40.3\\
\rowcolor{green!10}
 & & & \textbf{47.8} & \textbf{45.3}& \textbf{42.9}& \textbf{41.9}& \textbf{42.4}\\
\rowcolor{green!10}
\multirow{-2}{*}
{\textbf{QKVQA\textsuperscript{\small$\ddagger$} (ours)}} & \multirow{-2}{*}{LLaMA-3.1-8B} & \multirow{-2}{*}{$\checkmark$} & \textbf{+1.3\%} & \textbf{+1.2\%} & \textbf{+1.5\%}& \textbf{+1.3\%}& \textbf{+1.6\%}\\
\midrule

EchoSight~\cite{yan2024echosight} & LLaMA-3-8B & $\times$ & 41.8 & -- & -- & -- & 31.3 \\
OMGM\textsuperscript{\small$\dagger$}~\cite{yang2025omgm} & LLaMA-3-8B  & $\times$ & \underline{49.7} & \underline{45.7} & \underline{34.2} & \underline{32.1} & \underline{33.1} \\
\rowcolor{green!10}
 & & & \textbf{52.8} & \textbf{48.5} & \textbf{35.1} & \textbf{33.8} & \textbf{34.5} \\
\rowcolor{green!10}
\multirow{-2}{*}{\textbf{QKVQA (ours)}} & \multirow{-2}{*}{LLaMA-3-8B} & \multirow{-2}{*}{$\times$} & \textbf{+3.1\%} & \textbf{+2.8\%} & \textbf{+0.9\%} & \textbf{+1.7\%} & \textbf{+1.4\%} \\
\midrule
OMGM\textsuperscript{\small$\dagger$}~\cite{yang2025omgm} & Qwen2.5-VL-7B  & $\times$ & \underline{48.8} & \underline{44.8} & \underline{35.4} & \underline{32.9} & \underline{34.1} \\
\rowcolor{green!10}
 & & & \textbf{51.8} & \textbf{48.0} & \textbf{36.3} & \textbf{36.3} & \textbf{36.3} \\
\rowcolor{green!10}
\multirow{-2}{*}{\textbf{QKVQA (ours)}} & \multirow{-2}{*}{Qwen2.5-VL-7B} & \multirow{-2}{*}{$\times$} & \textbf{+3.0\%} & \textbf{+3.2\%} & \textbf{+0.9\%} & \textbf{+3.4\%} & \textbf{+2.2\%} \\

\bottomrule
\end{tabular}
}
}

\vspace{-0.2cm}
\end{table*}

\begin{table*}[t]
\caption{VQA performance, average context length (in tokens), and computational time on the E-VQA and InfoSeek datasets.}
\vspace{-0.2cm}
\label{tab:ablation_stepbystep_latency}
\setlength{\tabcolsep}{1.5mm}
\centering
\begin{tabular}{c  c  c c c c c c}
\toprule
\textbf{Dataset} & \textbf{Method} & \textbf{Retrieval Time} & \textbf{Filtering Time} & \textbf{Inference Time} & \textbf{Total Time} & \textbf{Accuracy} & \textbf{Context Length}  \\
\midrule
\multirow{3}{*}{E-VQA} 
  & MLLM-based & 0.08 & 9.77 & 1.93 & 11.78 & 37.7 & 7411 \\
   & OMGM & 0.08 & \textbf{1.58} & 1.73 & 3.39 & 44.8 & 279 \\
 & \textbf{\cellcolor{green!10}QKVQA} & \textbf{\cellcolor{green!10}0.08}  & \cellcolor{green!10}1.67 & \textbf{\cellcolor{green!10}1.56} & \textbf{\cellcolor{green!10}3.31} & \textbf{\cellcolor{green!10}48.0} & \textbf{\cellcolor{green!10}246} \\

\midrule
\multirow{2}{*}{InfoSeek} 
 & OMGM & 0.09 & \textbf{2.07} & 0.39 & \textbf{2.55} & 34.1 & 363 \\
 & \textbf{\cellcolor{green!10}QKVQA} & \textbf{\cellcolor{green!10}0.09}  & \cellcolor{green!10}2.30 & \textbf{\cellcolor{green!10}0.38} & \cellcolor{green!10}2.77 & \textbf{\cellcolor{green!10}36.3} & \textbf{\cellcolor{green!10}334} \\
\bottomrule
\end{tabular}
\vspace{-3mm}
\end{table*}

\subsection{Main Results}
\subsubsection{VQA Result}
Table \ref{tab:vqa_main_results} presents the performance comparison between QKVQA and baseline models on the E-VQA and InfoSeek datasets. Models are evaluated under three configurations: (1) Base models without any filter augmentation or fine-tuning, including LLaMA-3-8B~\cite{dubey2024llama}, LLaMA-3.1-8B~\cite{dubey2024llama}, GPT-4~\cite{achiam2023gpt}, Qwen2.5-VL-7B~\cite{bai2025qwen2}, and GPT-4V~\cite{achiam2023gpt}; (2) Without generator fine-tuning, where EchoSight~\cite{yan2024echosight} and OMGM~\cite{yang2025omgm} are compared; (3) With generator fine-tuning, which includes VLM-PRF~\cite{hong2025knowledge}, mKG-RAG~\cite{yuan2025mkg}, ReflectiVA~\cite{cocchi2025augmenting}, and OMGM as comparison methods. To ensure a fair comparison in the generator fine-tuning setting, both our method and OMGM adopt the publicly released fine-tuned generator from ReflectiVA. Among the selected baselines, VLM-PRF, mKG-RAG, and ReflectiVA are filtering methods based on Multimodal Large Language Models (MLLMs), for which we select their reported results using the same LLM as the generator. In contrast, EchoSight and OMGM are typical non-MLLM filtering methods. Experimental results indicate that typical methods do not fall significantly behind MLLM-based approaches in terms of VQA accuracy, which can be attributed to their focus on a more limited information scope, thereby demonstrating better noise control.

QKVQA achieves the best performance across all settings and metrics. Under the setting with LLaMA-3.1-8B as the fine-tuned generator, it demonstrates notable improvements: QKVQA attains 45.3\% on E-VQA, surpassing OMGM by 1.2\%; on InfoSeek-Overall, it reaches 42.4\%, outperforming VLM-PRF (+1.6\%) and OMGM (+2.1\%). In the mode without generator fine-tuning using LLaMA-3-8B, QKVQA achieves 48.5\% on E-VQA and 34.5\% on InfoSeek-Overall, exceeding the previous best model OMGM by 2.8\% and 1.4\%, respectively. When employing Qwen2.5-VL-7B as the generator under the setting without generator fine-tuning, QKVQA maintains its leading advantage: it achieves 48.0\% on E-VQA (an improvement of 3.2\% over OMGM) and 36.3\% on InfoSeek-Overall (an improvement of 2.2\% over OMGM). The consistent performance gains across diverse generator architectures demonstrate the strong generalizability of our filtering approach.

\subsubsection{Analysis of Context Length and Time Consumption}
Table~\ref{tab:ablation_stepbystep_latency} compares the time consumption at each stage, answer accuracy, and average context length (in tokens) between our method and OMGM on the E-VQA and InfoSeek datasets, both using Qwen2.5-VL-7B as the generator. On the E-VQA dataset, the MLLM-based method employs Qwen2.5-VL-7B for both filtering and generation. It first filters the top-5 retrieved articles~\cite{cocchi2025augmenting,hong2025knowledge} and then merges the selected information for generation (the corresponding prompt can be found in the Appendix). As shown in the table, this method incurs significant latency during the filtering stage (9.77\,s), severely limiting its practicality. Moreover, due to the lack of fine-grained noise control, its VQA accuracy (37.7\%) falls below both OMGM (44.8\%) and our method (48.0\%). Compared to OMGM, QKVQA's chunk-based CDA handles a larger number of processing units, resulting in slightly inferior filtering time, but outperforms OMGM in both generation time and total time, while achieving higher accuracy with shorter context length.

On the InfoSeek dataset, due to the larger number of chunks, our method slightly lags behind OMGM in filtering time and total time, but remains within a comparable range. Meanwhile, we still achieve superior accuracy with shorter context and generation time.

\begin{figure*}[ht!]
% 第一行
\begin{minipage}[b]{0.325\linewidth}
\begin{minipage}{0.44\linewidth}
\includegraphics[width=1.\linewidth]{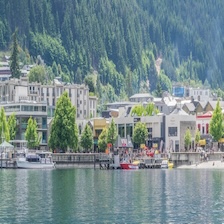}
\end{minipage}
\hfill
\begin{minipage}{0.54\linewidth}
\scriptsize{
\textbf{Question:} Which river flows out from this lake?\vspace{0.11cm}\\
\textbf{OMGM:} ``\#Wiki Article: Queenstown, New Zealand...''\\No river flows out of this lake. \textcolor{red}{\XSolidBold} \\
\textbf{QKVQA:} ``\#Wiki Article: Lake Wakatipu...The lake is drained by the Kawarau River...''\\Kawarau River. \textcolor[HTML]{00b050}{\CheckmarkBold}
}
\end{minipage}
\end{minipage}
\hfill
\begin{minipage}[b]{0.325\linewidth}
\begin{minipage}{0.44\linewidth}
\includegraphics[width=1.\linewidth]{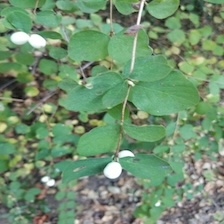}
\end{minipage}
\hfill
\begin{minipage}{0.54\linewidth}
\scriptsize{
\textbf{Question:} How many varieties of this plant are there?\vspace{0.05cm}\\
\textbf{OMGM:} ``\#Wiki Article: Symphoricarpos rotundifolius...''\\Not provided.\textcolor{red}{\XSolidBold} \\
\textbf{QKVQA:} ``\#Wiki Article: Symphoricarpos albus...There are two varieties...''\\Two varieties. \textcolor[HTML]{00b050}{\CheckmarkBold}
}
\end{minipage}
\end{minipage}
\hfill
\begin{minipage}[b]{0.325\linewidth}
\begin{minipage}{0.44\linewidth}
\includegraphics[width=1.\linewidth]{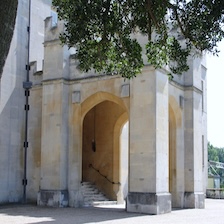}
\end{minipage}
\hfill
\begin{minipage}{0.54\linewidth}
\scriptsize{
\textbf{Question:} When was this house built?\vspace{0.05cm}\\
\textbf{OMGM:} ``\#Wiki Article: Syon House \#Section Title: Syon House...''\\
In the 1760s. \textcolor{red}{\XSolidBold} \\
\textbf{QKVQA:} ``\#Wiki Article: Syon House \#Section Title: History...\#Section Title: Architecture...was erected in 1547...''\\
1547.  \textcolor[HTML]{00b050}{\CheckmarkBold}
}
\end{minipage}
\end{minipage}

\vspace{0.06cm}

% 第二行
\begin{minipage}[b]{0.325\linewidth}
\begin{minipage}{0.44\linewidth}
\includegraphics[width=1.\linewidth]{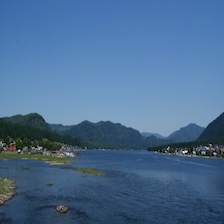}
\end{minipage}
\hfill
\begin{minipage}{0.54\linewidth}
\scriptsize{
\textbf{Question:} What is the length in kilometre is the lake in the image?\vspace{0.11cm}\\
\textbf{OMGM:} ``\#Wiki Article: Mana (river)...''\\475. \textcolor{red}{\XSolidBold} \\
\textbf{QKVQA:} ``\#Wiki Article: Lake Teletskoye...the lake is 78 km...\#Wiki Article: Mana (river)...''\\78. \textcolor[HTML]{00b050}{\CheckmarkBold}
}
\end{minipage}
\end{minipage}
\hfill
\begin{minipage}[b]{0.325\linewidth}
\begin{minipage}{0.44\linewidth}
\includegraphics[width=1.\linewidth]{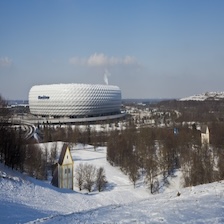}
\end{minipage}
\hfill
\begin{minipage}{0.54\linewidth}
\scriptsize{
\textbf{Question:} Who is the owner of this building?\vspace{0.06cm}\\
\textbf{OMGM:} ``\#Wiki Article: Architecture of Munich...''\\Bavaria. \textcolor{red}{\XSolidBold} \\
\textbf{QKVQA:}  ``\#Wiki Article: Allianz Arena...Bayern Munich took over all the shares and owns 100 per cent of the Allianz Arena...''\\Bayern Munich. \textcolor[HTML]{00b050}{\CheckmarkBold}
}
\end{minipage}
\end{minipage}
\hfill
\begin{minipage}[b]{0.325\linewidth}
\begin{minipage}{0.44\linewidth}
\includegraphics[width=1.\linewidth]{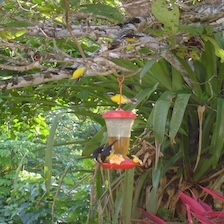}
\end{minipage}
\hfill
\begin{minipage}{0.54\linewidth}
\scriptsize{
\textbf{Question:} What is the closest parent taxonomy of this bird?\vspace{0.05cm}\\
\textbf{OMGM:} ``\#Wiki Article: Bananaquit \#Section Title: Bananaquit...''\\Thraupidae.\textcolor{red}{\XSolidBold} \\
\textbf{QKVQA:}``\#Wiki Article: Bananaquit \#Section Title: Taxonomy...as the only member of the genus Coereba...''\\Coereba.\textcolor[HTML]{00b050}{\CheckmarkBold}
}
\end{minipage}
\end{minipage}

\vspace{-2mm}
\caption{Qualitative comparison between OMGM and our proposed QKVQA method on E-VQA (top row) and InfoSeek (bottom row) image-question pairs. For OMGM in each row: the first two cases show article selection errors, while the last one exhibits intra-article information selection errors. For each result, the content within quotes represents the final context filtered by the corresponding method. Our QKVQA method successfully alleviates these issues and generates more accurate answers.}
\label{fig:qualitative_comparison}
\vspace{-3mm}
\end{figure*}

\begin{table}[h]
\caption{Performance comparison of QKVQA and OMGM on three key error-related metrics. R1, R1-C, and NR-C are higher the better.}
\vspace{-2mm}
\label{tab:two_error}
\setlength{\tabcolsep}{2mm}
\centering
\begin{tabular}{c c c c c}
\toprule
\textbf{Dataset} & \textbf{Method} & \textbf{R1} & \textbf{R1-C} & \textbf{NR-C} \\
\midrule
\multirow{2}{*}{E-VQA} 
& OMGM & 42.5 & 85.6 & 12.4 \\
& \textbf{\cellcolor{green!10}QKVQA} & \textbf{\cellcolor{green!10}43.7} & \textbf{\cellcolor{green!10}87.1} & \textbf{\cellcolor{green!10}13.8} \\
\cmidrule{1-5}
\multirow{2}{*}{InfoSeek}
& OMGM & 63.4 & 45.1 & 4.9 \\
& \textbf{\cellcolor{green!10}QKVQA} & \textbf{\cellcolor{green!10}63.7} & \textbf{\cellcolor{green!10}47.7} & \textbf{\cellcolor{green!10}6.0} \\
\bottomrule
\end{tabular}
\vspace{-3mm}
\end{table}

\subsubsection{Analysis of Two Key Error Types}

To investigate the improvement mechanism of QKVQA, we analyze it based on two core errors: article selection error and intra-article information selection error. Figure~\ref{fig:qualitative_comparison} presents qualitative comparisons using the Qwen2.5-VL-7B model as the generator on the E-VQA and InfoSeek datasets. Based on whether the correct article is ranked first and whether the final answer is correct, we define three metrics: R1 (Rank-1) as the proportion of samples where the correct article is ranked first; R1-C (Rank-1 with Correct Answer) as among samples where the correct article is ranked first, the proportion where the answer is correct; and NR-C (Non-Rank-1 with Correct Answer) as the proportion of samples where the correct article is not ranked first but the answer is correct.

The improvement in R1 demonstrates that QFF more accurately ranks the correct article first, mitigating article-level selection errors. The improvement in R1-C reflects that CDA expands the selection range within the article, enabling more precise intra-article information filtering and reducing errors inside the correct article, as shown in the last example of the first row in Figure~\ref{fig:qualitative_comparison}. Meanwhile, the improvement in NR-C benefits from CDA expanding the selection range of articles, enabling the correct article and its evidence sentences to be included in the final context. Although CDA may introduce irrelevant articles during the expansion of the selection range, as shown in the first example of the second row in Figure~\ref{fig:qualitative_comparison}, the generator can still select the correct document information for answering based on content ranking, the relevance of article information to the question, and the visual information from the image associated with the question. In summary, QKVQA achieves more precise selection through enhancing the filter's ability to focus on parts relevant to the question during filtering and expands the knowledge selection range, incorporating evidence from the correct article into the final context while compressing context length, thereby effectively mitigating the two types of errors.

\begin{table}[h]
\caption{The effect of QFF on the retrieval results. w/o. QFF denotes QKVQA without the QFF component.}
\vspace{-2mm}

\label{tab:ablation_retrieval_scope}
% \small
\setlength{\tabcolsep}{1mm}
\centering
\begin{tabular}{c c ccccc}
\toprule
\textbf{Dataset} & \textbf{Method} & \textbf{R@1} & \textbf{R@2} & \textbf{R@3} & \textbf{R@5} & \textbf{R@10} \\
\midrule
\multirow{3}{*}{E-VQA} 
& OMGM & 42.5 & 49.6 & 52.8 & 55.7 & 58.0 \\
& w/o. QFF & 18.7 & 28.5 & 33.8 & 41.2 & 49.6\\
                        & \textbf{\cellcolor{green!10}QKVQA}  & \textbf{\cellcolor{green!10}43.7} & 
                        \textbf{\cellcolor{green!10}50.1} &
                        \textbf{\cellcolor{green!10}53.5} &
                        \textbf{\cellcolor{green!10}56.0} & \textbf{\cellcolor{green!10}58.2}\\
\midrule
\multirow{3}{*}{InfoSeek} 
& OMGM & 63.4 & 73.0 & 76.9 & 80.4 & 83.3  \\
& w/o. QFF & 52.2 & 62.8 & 68.0 & 73.7 & 79.8 \\
                          & \textbf{\cellcolor{green!10}QKVQA}   & \textbf{\cellcolor{green!10}63.7} & 
                          \textbf{\cellcolor{green!10}73.1} &
                          \textbf{\cellcolor{green!10}77.1} &
                          \textbf{\cellcolor{green!10}80.4} & \textbf{\cellcolor{green!10}83.5} \\
\bottomrule
\end{tabular}

\vspace{-3mm}
\end{table}

\subsubsection{Retrieval Result}
\label{4.3.4}
We evaluate QFF on the E-VQA and InfoSeek datasets with a retrieval scope of 20 articles (Table~\ref{tab:ablation_retrieval_scope}). The results show consistent recall improvements across all metrics compared to both the baseline (QKVQA without QFF) and the existing method OMGM.

On E-VQA, QFF outperforms OMGM across all recall metrics, with particularly substantial gains over the baseline. On InfoSeek, despite being evaluated in a cross-domain setting without training on this dataset, QFF achieves performance slightly better than OMGM and significantly surpasses the baseline across all metrics. These results confirm that QFF effectively guides the filtering system to focus on relevant sections, achieving performance comparable to or better than OMGM while substantially improving upon the baseline.

\begin{table*}[t]
\caption{Ablation study results on E-VQA and InfoSeek. CL denotes average context length (in tokens).}
\vspace{-2mm}

\label{tab:ablation}
% \vspace{-0.3cm}
  \centering
  \setlength{\tabcolsep}{0.3em}
  \begin{tabular}{cccccccccccc}
   \toprule
   \multicolumn{2}{c}{\textbf{QFF}} & \multicolumn{3}{c}{\textbf{CDA}} & \multicolumn{3}{c}{\textbf{E-VQA}} & \multicolumn{4}{c}{\textbf{InfoSeek}} \\
   \cmidrule(r){1-2} \cmidrule(lr){3-5} \cmidrule(lr){6-8} \cmidrule(l){9-12}
   Fine-tuning & Question-focused & Chunk-based & Cross-chunk & Cross‑article & Single-Hop & All & CL & Unsee-Q & Unseen-E & Overall & CL\\
    $\times$ & $\times$ & $\times$ & $\times$ & $\times$  & 36.17 & 34.16 & 205 & 17.74 & 17.23 & 17.48 & 103 \\
    $\checkmark$ & $\times$ & $\times$ & $\times$ & $\times$ & 47.83 & 44.19 & 294 & 34.92 & 30.83 & 32.75 & 375 \\
    $\checkmark$ & $\checkmark$ & $\times$ & $\times$ & $\times$  & 50.59 & 46.77 & 281 & 36.22 & 34.48 & 35.33 & 408 \\
    $\checkmark$ & $\checkmark$ & $\checkmark$ & $\times$ & $\times$  & 48.17 & 44.59 & 55 & 33.07 & 33.06 & 33.06 & 86 \\
    $\checkmark$ & $\checkmark$ & $\checkmark$ & $\checkmark$ & $\times$  & 50.95 & 47.15 & 213 & 35.49 & 35.59 & 35.54 & 294 \\
    \rowcolor{green!10}
    $\checkmark$ & $\checkmark$ & $\checkmark$ & $\checkmark$ & $\checkmark$ & 51.81 & 48.02 & 246 & 36.26 & 36.31 & 36.28 & 334 \\   
   \bottomrule
  \end{tabular}
  % \vspace{-0.1cm}

% \vspace{-0.8cm}
\end{table*}

\subsection{Ablation Study}
\subsubsection{Effect of QFF and CDA}

We analyze the contribution of each component through ablation studies (Table \ref{tab:ablation}). Using the Qwen2.5-VL-7B model, we progressively incorporate modules from a baseline. Before enabling the CDA module, we select the top-1 section from the top-1 article using QFF and BGE-Reranker-v2-m3~\cite{chen-etal-2024-m3}. We observe that without basic fine-tuning, QFF tends to prefer shorter sections, which mostly lack evidence, resulting in low accuracy.

Basic fine-tuning of QFF alone yields significant performance gains on both datasets (E-VQA from 34.16\% to 44.19\%; InfoSeek Overall from 17.48\% to 32.75\%), demonstrating its necessity. Fine-tuning with the question-focused mechanism further improves results (E-VQA to 46.77\%; InfoSeek Overall to 35.33\%), confirming that QFF effectively guides article filtering by focusing on question semantics.

We then gradually enable CDA components. When selecting only the first chunk of the top-ranked article, performance drops due to substantial context discrepancy (55 vs. 281 tokens on E-VQA; 86 vs. 408 tokens on InfoSeek). However, when selecting a preset number of chunks from the top-1 article, despite using shorter context (213 vs. 281 tokens on E-VQA; 294 vs. 408 tokens on InfoSeek), accuracy recovers to 47.15\% on E-VQA and 35.54\% on InfoSeek, exceeding that of selecting a single section (46.77\% and 35.33\%, respectively). This demonstrates that fine-grained chunk selection with expanded knowledge scope enables effective and concise information filtering. Finally, incorporating dynamic cross-article selection to form the complete QKVQA achieves the best performance on both datasets, reaching 48.02\% on E-VQA and 36.28\% on InfoSeek, with token counts of 246 and 334 respectively, still lower than the single-section approach (281 and 408 tokens).

\subsubsection{Effect of different hyper-parameters’ value in QFF and CDA}
\label{4.4.2}
\begin{figure}[h]
  \centering
  \includegraphics[width=\linewidth]{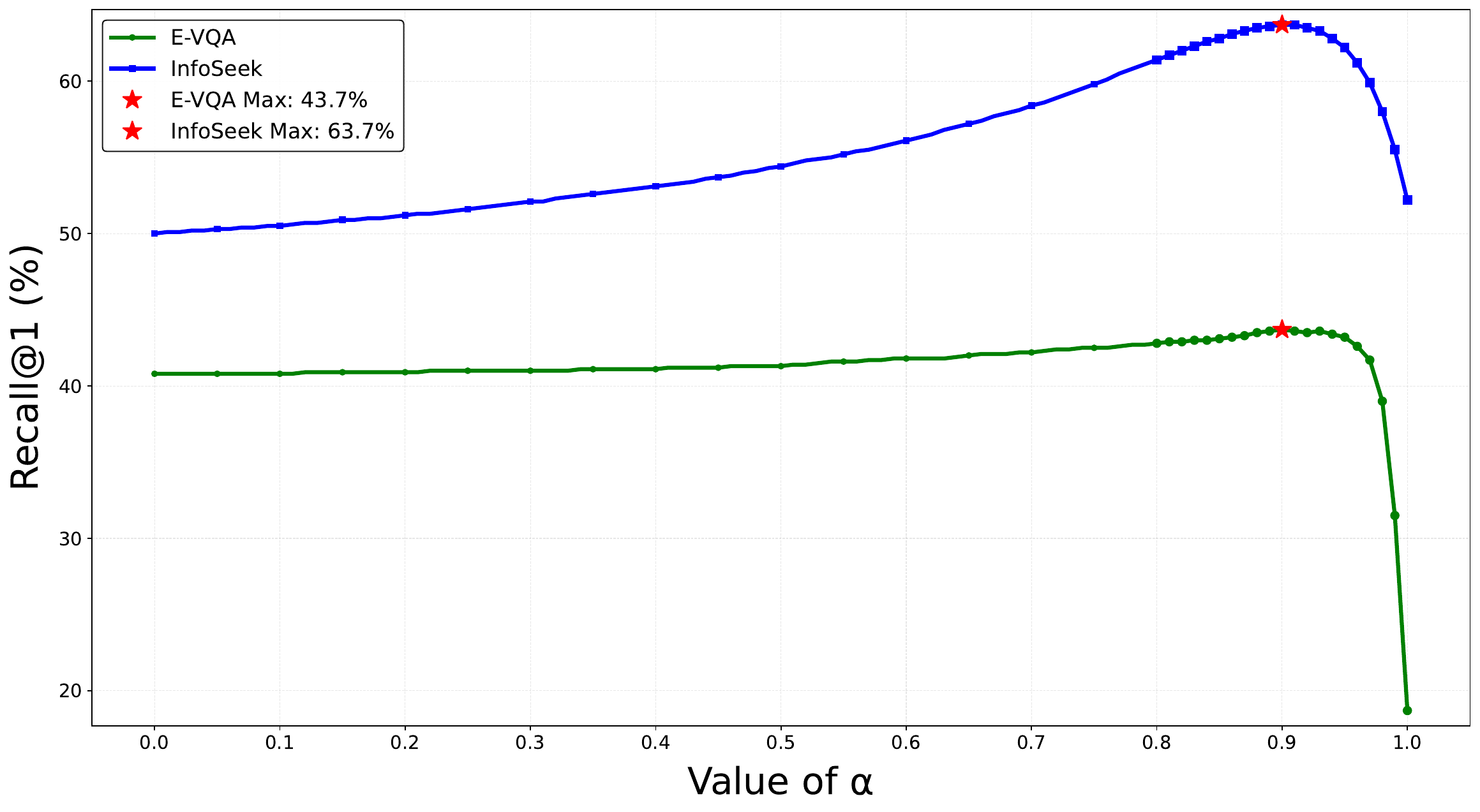}
  \caption{The variation of Recall@1 with the change of hyper-parameter $\alpha$ on the E-VQA and InfoSeek datasets.}
  \label{fig:alpha_lambda}
\end{figure}

In the filtering process of QFF, as shown in Equation~\ref{a}, we set $\alpha$ as the weight to integrate the coarse-grained similarity obtained from the initial article search with the multimodal fusion similarity, enabling comprehensive similarity computation. In our experiments with $\alpha$, we varied its value from $0$ to $1$ and observed the resulting changes in article recall@1, as illustrated on the left side of Figure~\ref{fig:alpha_lambda}. The results demonstrate that the fusion of similarities has a highly positive impact on reranking performance. Optimal retrieval performance is achieved when the coarse-grained retrieval similarity is mixed at a high proportion ($0.9$), indicating that the multimodal fusion similarity primarily refines the coarse-grained article similarity obtained from the initial retrieval.

We conducted experimental analysis on other hyperparameters involved in our method. All experiments were performed on a randomly sampled 1K subset of the InfoSeek dataset using Qwen2.5-VL-7B as the generator, with results shown in Table~\ref{tab:merged_hyperparameters}. On the InfoSeek dataset, we ultimately selected $U=2$, threshold $0.01$, and chunk size $90$ as the base configuration. When varying a specific hyperparameter, the remaining parameters were kept fixed at this configuration.

In the experiments adjusting $(K_1, K_2)$ and chunk length, we observed that under our proposed filtering framework, model accuracy exhibits a positive correlation with the final filtered context length. Experiments on $U$ and threshold $\theta$ demonstrate that indiscriminately expanding the candidate range (e.g., omitting $U$ or $\theta$) leads to performance degradation. This further validates the importance of fine-grained noise filtering while expanding the information scope.

\begin{table}[h]
\caption{VQA performance comparison under different hyper-parameter configurations. CL denotes average context length (in tokens).}
\label{tab:merged_hyperparameters}
\centering
\setlength{\tabcolsep}{2mm}
\begin{tabular}{c c c c}
\toprule
\textbf{Hyper-parameter} & \textbf{Values} & \textbf{Accuracy} & \textbf{CL} \\
\midrule
\multirow{4}{*}{$U$} 
 & 1      & 37.20 & 292 \\
 & 2      & 37.53 & 325 \\
 & 3      & 37.19 & 334 \\
 & None   & 37.26 & 337 \\
\midrule
\multirow{5}{*}{$\theta$} 
 & 0      & 37.20 & 292 \\
 & 0.01   & 37.53 & 325 \\
 & 0.02   & 37.76 & 349 \\
 & 0.03   & 36.95 & 364 \\
 & None   & 34.03 & 440 \\
\midrule
\multirow{4}{*}{Chunk Length} 
 & 45    & 37.13 & 237 \\
 & 90    & 37.53 & 325 \\
 & 180    & 39.43 & 614 \\
 & 360   & 39.61 & 959 \\
\midrule
\multirow{4}{*}{($K_1$, $K_2$)} 
 & (2,1)  & 35.17 & 166 \\
 & (4,2)  & 37.53 & 325 \\
 & (4,4)  & 38.00 & 356 \\
 & (8,4)  & 39.58 & 628 \\
\bottomrule
\end{tabular}
\vspace{-2mm}
\end{table}

\section{Conclusion}
We propose QKVQA, a question-focused filtering framework for Knowledge-based VQA (KB-VQA). The framework is designed to enhance the ability to focus on parts relevant to the question during candidate section encoding, expanding the scope of knowledge selection while reducing context length. This mitigates article selection errors and intra-article information selection errors, ultimately improving KB-VQA accuracy. It comprises two core modules: a Question-Focused Filter (QFF) that injects question semantics into candidate section encoding, and a Chunk-based Dynamic Cross-Article Selection module (CDA) that performs fine-grained cross-article filtering. Experiments on E-VQA and InfoSeek show that QKVQA outperforms state-of-the-art methods with shorter context and inference time comparable to the optimal approach, validating its effectiveness and offering insights for context refinement.

\clearpage
\bibliographystyle{ACM-Reference-Format}
\bibliography{samples/main}

% \appendix

\section{Prompts Details in QKVQA}

\subsection{LLM for E-VQA}
\begin{tcolorbox}[colback=white,colframe=black!75,title=System Prompt,fonttitle=\bfseries, breakable, enhanced]
Answer the encyclopedic question about the given image. Don't mention the visual content of image in your output. Directly output the answer of the question according to the context.\\
You are a helpful assistant for answering encyclopedic questions.\\

If the context does not contain the information required to answer the question, you should answer the question using internal model knowledge.
\end{tcolorbox}

\begin{tcolorbox}[colback=white,colframe=black!50,title=User Prompt, breakable, enhanced]
- Context: \{ \textit{Chunks} \}\\
- Question: \{ \textit{Textual question} \}\\
The answer is:
\end{tcolorbox}

\subsection{VLM for E-VQA}
\begin{tcolorbox}[colback=white,colframe=black!75,title=System Prompt,fonttitle=\bfseries, breakable, enhanced]
Answer the encyclopedic question about the given image. Don't mention the visual content of image in your output. Directly output the answer of the question according to the context.\\
If the context does not contain the information required to answer the question, you should answer the question using internal model knowledge.
\end{tcolorbox}

\begin{tcolorbox}[colback=white,colframe=black!50,title=User Prompt, breakable, enhanced]
- Context: \{ \textit{Chunks} \}\\
- Question: \{ \textit{Textual question} \}\\
\{ \textit{Query Image} \}\\
The answer is:
\end{tcolorbox}

\subsection{LLM for InfoSeek}
\begin{tcolorbox}[colback=white,colframe=black!75,title=System Prompt,fonttitle=\bfseries, breakable, enhanced]
Answer the encyclopedic question about the given image. Don't mention the visual content of image in your output. Directly output the answer of the question according to the context.\\
You are a helpful assistant for answering encyclopedic questions. Do not answer anything else.\\
If you need to answer questions about numbers or time, please output the corresponding numerical format directly. If the context does not contain the information required to answer the question, you should answer the question using internal model knowledge.\\
There is an example:\\
- Context: \# Wiki Article: Dolomites\\
\#\#Section Title: Dolomites\\
The Dolomites, also known as the Dolomite Mountains, Dolomite Alps or Dolomitic Alps, are a mountain range located in northeastern Italy. The Dolomites are located in the regions of Veneto, Trentino-Alto Adige/Südtirol and Friuli Venezia Giulia, covering an area shared between the provinces of Belluno, Vicenza, Verona, Trentino, South Tyrol, Udine and Pordenone.\\
- Question: Which city or region does this mountain locate in?\\
Just answer the questions, no explanations needed. Short answer is: Province of Belluno
\end{tcolorbox}

\begin{tcolorbox}[colback=white,colframe=black!50,title=User Prompt, breakable, enhanced]
- Context: \{ \textit{Chunks} \}\\
- Question: \{ \textit{Textual question} \}\\
Just answer the questions, no explanations needed. Short answer is:
\end{tcolorbox}

\subsection{VLM for InfoSeek}
\begin{tcolorbox}[colback=white,colframe=black!75,title=System Prompt,fonttitle=\bfseries, breakable, enhanced]
Answer the encyclopedic question about the given image. Don't mention the visual content of image in your output. Directly output the answer of the question according to the context.\\
If you need to answer questions about numbers or time, please output the corresponding numerical format directly. If the context does not contain the information required to answer the question, you should answer the question using internal model knowledge.\\
There is an example:\\
- Context: \# Wiki Article: Dolomites\\
\#\#Section Title: Dolomites\\
The Dolomites, also known as the Dolomite Mountains, Dolomite Alps or Dolomitic Alps, are a mountain range located in northeastern Italy. The Dolomites are located in the regions of Veneto, Trentino-Alto Adige/Südtirol and Friuli Venezia Giulia, covering an area shared between the provinces of Belluno, Vicenza, Verona, Trentino, South Tyrol, Udine and Pordenone.\\
- Question: Which city or region does this mountain locate in?\\
Just answer the questions, no explanations needed. Short answer is: Province of Belluno
\end{tcolorbox}

\begin{tcolorbox}[colback=white,colframe=black!50,title=User Prompt, breakable, enhanced]
- Context: \{ \textit{Chunks} \}\\
- Question: \{ \textit{Textual question} \}\\
\{ \textit{Query Image} \}\\
Just answer the questions, no explanations needed. Short answer is:
\end{tcolorbox}

\subsection{VLM for Filtering}
\begin{tcolorbox}[colback=white,colframe=black!75,title=System Prompt,fonttitle=\bfseries, breakable, enhanced]
You are an encyclopedia expert. Given the question, image and candidate article, filter out the information in the article that is relevant to answering the question. Do not directly answer the question.
\end{tcolorbox}

\begin{tcolorbox}[colback=white,colframe=black!50,title=User Prompt, breakable, enhanced]
\{ \textit{Query Image} \}\\
- Question: \{ \textit{Textual question} \}\\
\{ \textit{Candidate Image} \}\\
- Context: \{ \textit{Sections} \}

\end{tcolorbox}

% \clearpage  % 或 \newpage

\section{Dataset Details}

\noindent \textbf{Encyclopedic Visual Question Answering Dataset}\\
This dataset consists of approximately 221,000 question-answer pairs associated with 16,700 distinct fine-grained entities. Each entity is depicted by a maximum of five images. The fine-grained entities and their corresponding images are sourced from the iNaturalist 2021 dataset and the Google Landmarks Dataset V2. Furthermore, the dataset incorporates a structured knowledge base extracted from WikiWeb2M, encompassing 2 million Wikipedia articles accompanied by images, which provides supporting evidence for the answers. Questions within the dataset are categorized as single-hop or multi-hop based on the reasoning complexity required. The data splits include training, validation, and test sets, containing 1 million, 13,000, and 5,750 triplets, respectively (with 4,750 single-hop and 1,000 two-hop triplets in the test set). Our experiments focus exclusively on single-hop questions during the training phase.

To assess the effectiveness of our proposed retrieval-augmented large language model framework for visual question answering, we employ two standard evaluation metrics. For answer prediction quality, we adopt the BERT-based Evaluation Metric (BEM) score. This metric computes similarity between the predicted answer and the reference using a BERT model fine-tuned specifically for assessing answer similarity. Compared to conventional VQA evaluation metrics, this approach more robustly handles semantically correct answers that may not exactly match the annotated reference. Recall@K measures the retrieval performance by calculating the fraction of test instances where the ground-truth entity is present among the top-k retrieved results.

\noindent \textbf{InfoSeek Dataset}\\
This collection includes 1.3 million image-question-answer triplets corresponding to around 11k visual entities curated from the OVEN dataset. It contains 8,900 manually crafted visual information-seeking questions along with 1.3 million automatically generated ones. The triplets are partitioned into training, validation, and test subsets, comprising roughly 934k, 73k, and 348k samples, respectively. As ground-truth annotations for the test set are unavailable, all evaluations are performed on the validation split. Notably, both validation and test sets feature questions involving entities or queries not encountered during training. Additionally, the dataset provides access to a knowledge base of 6 million Wikipedia entities. Aligning with prior studies such as EchoSight, a subset of 100,000 entities is utilized, ensuring coverage of the 6,741 entities relevant to the training and validation questions. During the process of gathering images from Wikipedia pages for these entities, a minimal number of validation samples were found to correspond to correct entities lacking associated images. Following the methodology of EchoSight and OMGM, these samples are filtered out. Evaluations are thus conducted on the remaining 71,335 validation instances, representing 96.9\% of the original validation set, thereby minimizing any potential impact on the final results.

Following the specific evaluation protocol of InfoSeek, two distinct metrics are applied based on answer types. For questions requiring string-based answers, such as entity names, we report accuracy using the VQA Accuracy metric. This metric accommodates multiple valid answers by considering minor phrasing variations as correct. A model's response is judged correct if it matches any of the acceptable answers exactly. For questions expecting numerical answers, the Relaxed Accuracy metric is used, where a prediction is considered correct if it falls within a predefined tolerance range around the ground-truth value. For retrieval assessment, Recall@K is retained as the evaluation metric, consistent with the approach for the Encyclopedic VQA dataset.

Table~\ref{tab:dataset_stats} presents the statistics of the two datasets used in our experiments.

\begin{table}[H]
    \caption{Statistics of InfoSeek and E-VQA datasets used in our experiments. 
    }
    \label{tab:dataset_stats}
    \centering
    \resizebox{\columnwidth}{!}{%
    \small
    \begin{tabular}{l cc c cc}
    \toprule
     \multirow{2}{*}{\textbf{Datasets}} & \multicolumn{2}{c}{\textbf{\#Samples}} & & \multicolumn{2}{c}{\textbf{\#Articles}} \\
     \cmidrule{2-3} \cmidrule{5-6} 
    & Train & Test & & Train & Valid/Test \\
    \midrule
    InfoSeek   & -- & 71,335 & & --  & 100K     \\
    E-VQA      & 190k & 5,750 & & 2M & 2M      \\
    \bottomrule
    \end{tabular}
    }

\end{table}

\section{Qualitative Results}
\begin{figure*}[t]
% 第一行
\begin{minipage}[b]{0.325\linewidth}
\begin{minipage}{0.44\linewidth}
\includegraphics[width=1.\linewidth]{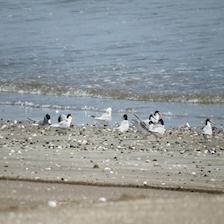}
\end{minipage}
\hfill
\begin{minipage}{0.54\linewidth}
\scriptsize{
\textbf{Question:} In which country or region does this animal live?
\vspace{0.018cm}\\
\textbf{OMGM:} ``\#Wiki Article: Yellow-billed tern...''\\South America.\textcolor{red}{\XSolidBold} \\
\textbf{QKVQA:} ``\#Wiki Article: Little tern...''\\Europe and Asia, South Africa and Australia. \textcolor[HTML]{00b050}{\CheckmarkBold}
}
\end{minipage}
\end{minipage}
\hfill
\begin{minipage}[b]{0.325\linewidth}
\begin{minipage}{0.44\linewidth}
\includegraphics[width=1.\linewidth]{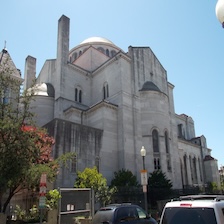}
\end{minipage}
\hfill
\begin{minipage}{0.54\linewidth}
\scriptsize{
\textbf{Question:} How many flags hang from the choir loft at this church?\vspace{0.05cm}\\
\textbf{OMGM:} ``\#Wiki Article: Saint Sophia Cathedral (Washington, D.C.)...''Not provided.
 \textcolor{red}{\XSolidBold} \\
\textbf{QKVQA:} ``\#Wiki Article: Shrine of the Sacred Heart...Several dozen flags hang from the church's choir loft...''Several dozen.
 \textcolor[HTML]{00b050}{\CheckmarkBold}
}
\end{minipage}
\end{minipage}
\hfill
\begin{minipage}[b]{0.325\linewidth}
\begin{minipage}{0.44\linewidth}
\includegraphics[width=1.\linewidth]{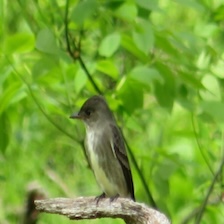}
\end{minipage}
\hfill
\begin{minipage}{0.54\linewidth}
\scriptsize{
\textbf{Question:} When was this bird classified as ``near threatened''?
\vspace{0.05cm}\\
\textbf{OMGM:} ``\#Wiki Article: Eastern wood pewee...''Not provided.
 \textcolor{red}{\XSolidBold} \\
\textbf{QKVQA:} ``\#Wiki Article: Olive-sided flycatcher...in 2016, ...is classified as a ``near threatened'' species...\#Wiki Article: Eastern wood pewee...''2016. \textcolor[HTML]{00b050}{\CheckmarkBold}
}
\end{minipage}
\end{minipage}

\vspace{0.2cm}

% 第二行
\begin{minipage}[b]{0.325\linewidth}
\begin{minipage}{0.44\linewidth}
\includegraphics[width=1.\linewidth]{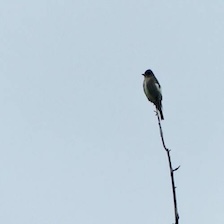}
\end{minipage}
\hfill
\begin{minipage}{0.54\linewidth}
\scriptsize{
\textbf{Question:} What type of flyer is this bird?\vspace{0.11cm}\\
\textbf{OMGM:} ``\#Wiki Article: Western wood pewee...''\\A type of flycatcher. \textcolor{red}{\XSolidBold} \\
\textbf{QKVQA:} ``\#Wiki Article: Olive-sided flycatcher...It is a very agile flyer...''\\Agile. \textcolor[HTML]{00b050}{\CheckmarkBold}
}
\end{minipage}
\end{minipage}
\hfill
\begin{minipage}[b]{0.325\linewidth}
\begin{minipage}{0.44\linewidth}
\includegraphics[width=1.\linewidth]{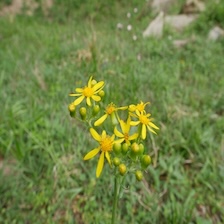}
\end{minipage}
\hfill
\begin{minipage}{0.54\linewidth}
\scriptsize{
\textbf{Question:} How many feet does this plant prefer in altitude?\vspace{0.06cm}\\
\textbf{OMGM:} ``\#Wiki Article: Senecio ovatus...''\\6,560 feet. \textcolor{red}{\XSolidBold} \\
\textbf{QKVQA:} ``\#Wiki Article: Senecio ampullaceus...to 800 meters (2,600 ft)...''\\2,600 feet. \textcolor[HTML]{00b050}{\CheckmarkBold}
}
\end{minipage}
\end{minipage}
\hfill
\begin{minipage}[b]{0.325\linewidth}
\begin{minipage}{0.44\linewidth}
\includegraphics[width=1.\linewidth]{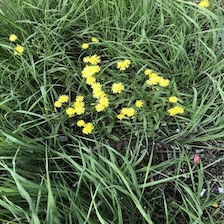}
\end{minipage}
\hfill
\begin{minipage}{0.54\linewidth}
\scriptsize{
\textbf{Question:} This plant prefers 3-4 cm of what in order to germinate?
\vspace{0.018cm}\\
\textbf{OMGM:} ``\#Wiki Article: Agoseris heterophylla...''\\
Rainfall. \textcolor{red}{\XSolidBold} \\
\textbf{QKVQA:} ``\#Wiki Article: Crepis tectorum...with an optimum depth of 3–4 cm...''\\
Depth. \textcolor[HTML]{00b050}{\CheckmarkBold}
}
\end{minipage}
\end{minipage}

\vspace{0.2cm}

% 第三行
\begin{minipage}[b]{0.325\linewidth}
\begin{minipage}{0.44\linewidth}
\includegraphics[width=1.\linewidth]{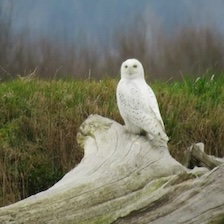}
\end{minipage}
\hfill
\begin{minipage}{0.54\linewidth}
\scriptsize{
\textbf{Question:} What part of the female of this bird is usually dark brown?\vspace{0.05cm}\\
\textbf{OMGM:} ``\#Wiki Article: Snowy owl \#Section Title: Snowy owl...''\\
Not provided. \textcolor{red}{\XSolidBold} \\

\textbf{QKVQA:} ``\#Wiki Article: Snowy owl \#Section Title: Description...on the crown and the underparts...''\\
The crown and the underparts. \textcolor[HTML]{00b050}{\CheckmarkBold}
}
\end{minipage}
\end{minipage}
\hfill
\begin{minipage}[b]{0.325\linewidth}
\begin{minipage}{0.44\linewidth}
\includegraphics[width=1.\linewidth]{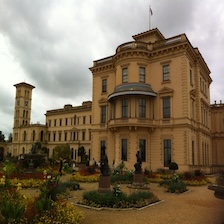}
\end{minipage}
\hfill
\begin{minipage}{0.54\linewidth}
\scriptsize{
\textbf{Question:} Who designed this house?\vspace{0.05cm}\\
\textbf{OMGM:} ``\#Wiki Article: Osborne House \#Section Title: Thomas Cubitt...''
Thomas Cubitt. \textcolor{red}{\XSolidBold} \\
\textbf{QKVQA:} ``\#Wiki Article: Osborne House \#Section Title: Royal retreat...Prince Albert designed the house himself...''
Prince Albert. \textcolor[HTML]{00b050}{\CheckmarkBold}
}
\end{minipage}
\end{minipage}
\hfill
\begin{minipage}[b]{0.325\linewidth}
\begin{minipage}{0.44\linewidth}
\includegraphics[width=1.\linewidth]{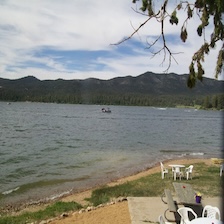}
\end{minipage}
\hfill
\begin{minipage}{0.54\linewidth}
\scriptsize{
\textbf{Question:} Which rivers flow into this reservoir?\vspace{0.05cm}\\
\textbf{OMGM:} ``\#Wiki Article: Big Bear Lake \#Section Title: History...''\\Not provided.  \textcolor{red}{\XSolidBold} \\
\textbf{QKVQA:} ``\#Wiki Article: Big Bear Lake \#Section Title: Geography...Bear Creek and Siberia Creek flow into the lake...''\\
Bear Creek and Siberia Creek. \textcolor[HTML]{00b050}{\CheckmarkBold}
}
\end{minipage}
\end{minipage}

\caption{Additional qualitative results on image-question pairs from Encyclopedic-VQA, where we compare the answers provided by QKVQA with those from OMGM. The first two rows demonstrate article selection errors by OMGM, while the last row shows an intra-article information selection error. For each result, the content within quotes represents the final context filtered by the corresponding method. Our QKVQA method provides correct answers in all cases.}
\label{fig:qualitatives_supp1}
\end{figure*}

\begin{figure*}[t]
% 第一行
\begin{minipage}[b]{0.325\linewidth}
\begin{minipage}{0.44\linewidth}
\includegraphics[width=1.\linewidth]{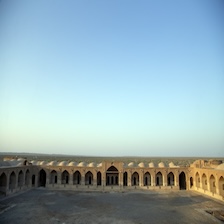}
\end{minipage}
\hfill
\begin{minipage}{0.54\linewidth}
\scriptsize{
\textbf{Question:} What is the length of this building in metre?\vspace{0.11cm}\\
\textbf{OMGM:} ``\#Wiki Article: Jameh Mosque of Natanz...''
37. \textcolor{red}{\XSolidBold} \\
\textbf{QKVQA:} ``\#Wiki Article: Dayr-e Gachin...The structure of the caravanserai is a 109 in 108 meter square...\#Wiki Article: Jameh Mosque of Natanz...''
109. \textcolor[HTML]{00b050}{\CheckmarkBold}
}
\end{minipage}
\end{minipage}
\hfill
\begin{minipage}[b]{0.325\linewidth}
\begin{minipage}{0.44\linewidth}
\includegraphics[width=1.\linewidth]{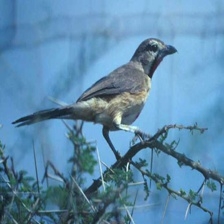}
\end{minipage}
\hfill
\begin{minipage}{0.54\linewidth}
\scriptsize{
\textbf{Question:} What is the closest upper taxonomy of this bird?\vspace{0.05cm}\\
\textbf{OMGM:} ``\#Wiki Article: Abyssinian waxbill...''\\
Estrildidae. \textcolor{red}{\XSolidBold} \\
\textbf{QKVQA:} ``\#Wiki Article: Bushshrike...have been defined as the superfamily Malaconotoidea...''\\
Malaconotoidea. \textcolor[HTML]{00b050}{\CheckmarkBold}
}
\end{minipage}
\end{minipage}
\hfill
\begin{minipage}[b]{0.325\linewidth}
\begin{minipage}{0.44\linewidth}
\includegraphics[width=1.\linewidth]{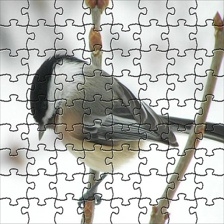}
\end{minipage}
\hfill
\begin{minipage}{0.54\linewidth}
\scriptsize{
\textbf{Question:} Who is the discoverer or inventor of this game?\vspace{0.05cm}\\
\textbf{OMGM:} ``\#Wiki Article: History of the London Underground...''\\
Not provided. \textcolor{red}{\XSolidBold} \\
\textbf{QKVQA:} ``\#Wiki Article: Jigsaw puzzle...John Spilsbury...is credited with commercialising jigsaw puzzles...''\\
John Spilsbury. \textcolor[HTML]{00b050}{\CheckmarkBold}
}
\end{minipage}
\end{minipage}

\vspace{0.2cm}

% 第二行
\begin{minipage}[b]{0.325\linewidth}
\begin{minipage}{0.44\linewidth}
\includegraphics[width=1.\linewidth]{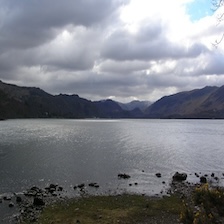}
\end{minipage}
\hfill
\begin{minipage}{0.54\linewidth}
\scriptsize{
\textbf{Question:} What place inflows lake?\vspace{0.05cm}\\
\textbf{OMGM:} ``\#Wiki Article: Ullswater...''\\
Aira Force waterfall. \textcolor{red}{\XSolidBold} \\
\textbf{QKVQA:} ``\#Wiki Article: Derwentwater...It is both fed and drained by the River Derwent...''\\
River Derwent. \textcolor[HTML]{00b050}{\CheckmarkBold}
}
\end{minipage}
\end{minipage}
\hfill
\begin{minipage}[b]{0.325\linewidth}
\begin{minipage}{0.44\linewidth}
\includegraphics[width=1.\linewidth]{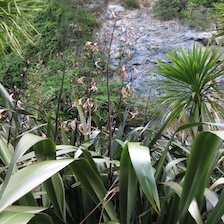}
\end{minipage}
\hfill
\begin{minipage}{0.54\linewidth}
\scriptsize{
\textbf{Question:} What is the closest parent taxonomy of this plant?\vspace{0.05cm}\\
\textbf{OMGM:} ``\#Wiki Article: Cordyline indivisa...''\\
Cordyline. \textcolor{red}{\XSolidBold} \\
\textbf{QKVQA:} ``\#Wiki Article: Phormium colensoi...''\\
Phormium. \textcolor[HTML]{00b050}{\CheckmarkBold}
}
\end{minipage}
\end{minipage}
\hfill
\begin{minipage}[b]{0.325\linewidth}
\begin{minipage}{0.44\linewidth}
\includegraphics[width=1.\linewidth]{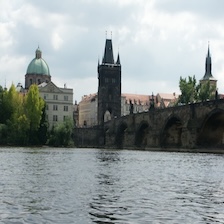}
\end{minipage}
\hfill
\begin{minipage}{0.54\linewidth}
\scriptsize{
\textbf{Question:} How many spans does this bridge have?\vspace{0.05cm}\\
\textbf{OMGM:} ``\#Wiki Article: Palacky Bridge...''\\
3. \textcolor{red}{\XSolidBold} \\
\textbf{QKVQA:} ``\#Wiki Article: Charles Bridge...it was built as a bow bridge with 16 arches...''\\
16. \textcolor[HTML]{00b050}{\CheckmarkBold}
}
\end{minipage}
\end{minipage}

\vspace{0.2cm}

% 第三行
\begin{minipage}[b]{0.325\linewidth}
\begin{minipage}{0.44\linewidth}
\includegraphics[width=1.\linewidth]{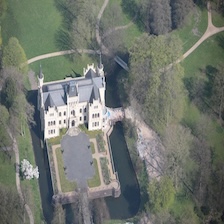}
\end{minipage}
\hfill
\begin{minipage}{0.54\linewidth}
\scriptsize{
\textbf{Question:} What country does this building belong to?\vspace{0.05cm}\\
\textbf{OMGM:} ``\#Wiki Article: Evenburg \#Section Title: History...''\\
Norway. \textcolor{red}{\XSolidBold} \\
\textbf{QKVQA:} ``\#Wiki Article: Evenburg \#Section Title: History...\#Section Title: Evenburg...in north Germany...''\\
Germany. \textcolor[HTML]{00b050}{\CheckmarkBold}
}
\end{minipage}
\end{minipage}
\hfill
\begin{minipage}[b]{0.325\linewidth}
\begin{minipage}{0.44\linewidth}
\includegraphics[width=1.\linewidth]{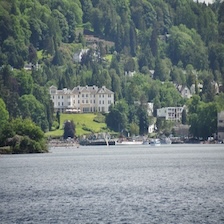}
\end{minipage}
\hfill
\begin{minipage}{0.54\linewidth}
\scriptsize{
\textbf{Question:} What is the vertical depth (in metre) of this lake?\vspace{0.05cm}\\
\textbf{OMGM:} ``\#Wiki Article: Windermere \#Section Title: Natural history...''
30. \textcolor{red}{\XSolidBold} \\
\textbf{QKVQA:} ``\#Wiki Article: Windermere \#Section Title: Geography...With a maximum depth of 66.7 m...''
66.7. \textcolor[HTML]{00b050}{\CheckmarkBold}
}
\end{minipage}
\end{minipage}
\hfill
\begin{minipage}[b]{0.325\linewidth}
\begin{minipage}{0.44\linewidth}
\includegraphics[width=1.\linewidth]{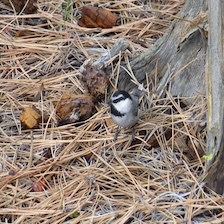}
\end{minipage}
\hfill
\begin{minipage}{0.54\linewidth}
\scriptsize{
\textbf{Question:} What is the closest parent taxonomy of this bird?\vspace{0.05cm}\\
\textbf{OMGM:} ``\#Wiki Article: Mountain chickadee \#Section Title: Breeding...''
Passeriformes. \textcolor{red}{\XSolidBold} \\
\textbf{QKVQA:} ``\#Wiki Article: Mountain chickadee \#Section Title: Taxonomy...that separating Poecile more adequately expresses these birds' relationships...''
Poecile. \textcolor[HTML]{00b050}{\CheckmarkBold}
}
\end{minipage}
\end{minipage}

\caption{Additional qualitative results on image-question pairs from InfoSeek, where we compare the answers provided by QKVQA with those from OMGM. The first two rows demonstrate article selection errors by OMGM, while the last row shows an intra-article information selection error. For each result, the content within quotes represents the final context filtered by the corresponding method. Our QKVQA method provides correct answers in all cases.}
\label{fig:qualitatives_supp2}
\end{figure*}

For a comprehensive evaluation of the proposed QKVQA model, we present additional qualitative results in Figures~\ref{fig:qualitatives_supp1} and ~\ref{fig:qualitatives_supp2}, comparing them with answers generated by OMGM. These results are based on sample image-question pairs from the Encyclopedic-VQA and InfoSeek datasets, respectively.

\end{document}